\documentclass{aa}
\usepackage{amsfonts}
\usepackage{amssymb}
\usepackage{amsgen}
\usepackage{amsmath}
\usepackage{amsmath}
\usepackage{epsfig}

\def\beb{\begin{thebibliography}{}}
\def\eeb{\end{thebibliography}}
\def\bei{\begin{itemize}}
\def\eei{\end{itemize}}
\def\bef{\begin{figure}}
\def\eef{\end{figure}}
\def\ben{\begin{enumerate}}
\def\een{\end{enumerate}}
\def\beq{\begin{equation}}
\def\eeq{\end{equation}}
\def\ber{\begin{eqnarray}}
\def\eer{\end{eqnarray}}

\newcommand{\lsim}{\raisebox{-0.3ex}{\mbox{$\stackrel{<}{_\sim} \,$}}}
\newcommand{\gsim}{\raisebox{-0.3ex}{\mbox{$\stackrel{>}{_\sim} \,$}}}

\begin{document}

\title{Searching for Pulsars in Close Binary Systems}

\author{S. Jouteux\inst{1} \and R. Ramachandran\inst{1}\inst{2} \and 
B.W. Stappers\inst{2} \and P.G. Jonker\inst{1} \and M. Van der Klis\inst{1}}
\offprints{ramach astro.uva.nl}
\institute{Sterrenkundig Instituut ``Anton Pannekoen'', Kruislaan 403, 1098
SJ Amsterdam, The Netherlands \and Stichting ASTRON, Postbus 2, 7990 AA
Dwingeloo, The Netherlands}

\abstract{We present a detailed mathematical analysis of the
Fourier response of binary pulsar signals whose frequencies are
modulated by circular orbital motion. The fluctuation power spectrum of
such signals is found to be $\nu_{\rm orb}$-periodic over a compact
frequency range, where $\nu_{\rm orb}$ denotes orbital frequency.
Subsequently, we consider a wide range of binary systems with circular
orbits and short orbital periods, and present a Partial Coherence
Recovery Technique for searching for binary millisecond X-ray and radio
pulsars. We use numerical simulations to investigate the detectability
of pulsars in such systems with $P_{\rm orb} \lsim 6$ hours, using
this technique and three widely used pulsar search methods. These
simulations demonstrate that the Partial Coherence Recovery Technique is
on average several times more sensitive at detecting pulsars in close
binary systems when the data span is more than 2 orbital periods. The
systems one may find using such a method can be used to improve the
constraints on the coalescence rate of compact objects and they also
represent those systems most likely to be detected with gravitational
wave detectors such as LISA. 
 }
\date{Received / Accepted}
\authorrunning{Jouteux et al.}
\titlerunning{Searching for pulsars in close binary systems}
\maketitle


\section{Introduction}

The importance of binarity and the connection with the low-mass X-ray
binaries (LMXBs) in the formation of millisecond pulsars (Alpar et
al. 1982; Radhakrishnan \& Srinivasan 1982) extends back to the
discovery of the first millisecond pulsar (Backer et al. 1982). In the
intervening years, a large number of binary millisecond radio pulsars
and X-ray pulsars have been discovered (for recent reviews see Kramer
et al. 2000). However, these tend to be either bright pulsars, members
of wide binaries or to have low-mass companions. The remaining systems
are more difficult to detect owing to large orbital accelerations
which cause, via the time-dependent Doppler effect, the apparent pulse
period to change significantly during long integrations. Compensating
completely for this smearing of the pulsed signal is extremely
computationally expensive as a large range of orbital parameters has
to be searched.

\noindent Numerous authors have developed and used techniques to
partially correct for this smearing, e.g.  Middleditch \& Priedhorsky
(1986); Anderson et al. (1990); Wood et al. (1991); Ransom (1999). In
1990 Anderson \emph{et\,al.}.  Anderson et al. (1990) used a constant
acceleration technique to successfully discover PSR 2127+11C in the
globular cluster M15. Despite the success of this survey, it was not
until recently that the true extent of the binary millisecond pulsar
population that could be uncovered by using such techniques became
apparent (Camilo et al. 2000; D'Amico et al. 2000; Ransom et
al. 2000). However a whole possible population of rapidly rotating
radio/X-ray pulsars in close binary systems still remains to be
investigated as even the current best search strategies are not very
sensitive to them. This population includes faint pulsars in similar
type binaries, pulsars in tighter binaries and those with more massive
companions.

\noindent Following the discovery of a large number of binary
millisecond pulsars in the globular cluster 47Tuc (Camilo et
al. 2000), Rasio, Pfahl \& Rappaport (2000) tried to model this
observed population.  They find that a large number of neutron star -
white dwarf binaries in even tighter orbits and possibly more massive
companions than those already discovered should exist in 47Tuc. The
existence of very short orbital period LMXBs such as 4U 1820-30
(Stella, White \& Priedhorsky 1987), and millisecond pulsars with
relatively massive companions, e.g. PSR B1744-24A (Lyne et al. 1990),
also indicate that very short period binary millisecond pulsars should
exist in globular clusters (Bisnovatyi-Kogan 1989; Ergma \& Fedarova
1991). Furthermore, multi-frequency radio imaging of globular clusters
has revealed several unidentified steep-spectrum radio sources whose
variable flux-densities and spectral indexes suggest radio pulsars
(Fruchter \& Goss 1990; 2000). Subsequent pulsed searches (e.g.  Lyne
et al. 1990; D'Amico 1993; Biggs et al. 1994) have detected
millisecond pulsars coincident with some of these point sources
supporting the idea that the remainder are also pulsars which are so
far undetected. Moreover, the number of pulsars detected in globular
clusters is far less than the expected population based on the
integrated flux density of the cores of some of the clusters. It has
been argued that even bright pulsars could have been missed by
previous searches due to Doppler smearing caused by the orbital motion
in close binary systems, even though their radio emission remains
detectable by interferometric observations. A further source of
possible pulsar candidates are the subset of the EGRET unidentified
$\gamma$-ray sources whose distribution above the Galactic disk has a
scale height of about 2 kpc (Grenier 2001). These sources have so far
evaded detection as radio pulsars and this may be because they are
members of close binaries.

\noindent The physics of gravitation in such extreme systems has
recently received growing attention because such binary systems will
be the most common known continuous sources of gravitational radiation
for gravitational wave detectors. The Laser Interferometer Space
Antenna (LISA) will be especially sensitive to compact neutron star
binaries with ultra-short orbital periods and large ($\gsim$ 0.5
M$_{\odot}$) companion masses (Benacquista, Portegies~Zwart \& Rasio
2000). Detecting and studying pulsed radio or X-ray emission from the
neutron stars in these systems will provide better constraints on the
binary parameters and therefore be of paramount importance in helping
to understand the gravitational wave emission (Dhurandhar \& vecchio
2000). Discovering such systems is further motivated by improving the
constraints on the coalescence rate of compact objects which remains
difficult to estimate. Therefore, both theoretical predictions and
observational evidence strongly encourage searching for highly
accelerated millisecond pulsars in globular clusters.

\noindent Thus motivated, we present here a detailed mathematical
analysis of the Fourier response of a radio/X-ray binary pulsar signal.
We then discuss the sensitivity of three widely used search techniques
applied to a range of binary systems. In Sec. \ref{sec:pcrt}, we present
a complete description of our Partial Coherence Recovery Technique,
which is based on the Phase Modulation Searching method first discussed
by Ransom (Ransom 1999). We also discuss the computational cost of the
method, how to estimate detection levels and how to derive all the
binary parameters directly from the Fourier signature of the binary.
Finally we compare the sensitivity of these different search techniques
and show that our method is not only computationally very efficient but
also significantly more sensitive to very tight binaries.

\medskip

\noindent A list of the main symbols we use in this paper accompanied by
a short description can be found in Appendix~\ref{appendix:table}.

\section{Pulsar signal} \label{sec:ps}

Consider a pulsar with a pulse frequency $\nu_{\rm psr}$, whose distance
from the observer $d(t)$ is changing with time $t$. The pulsar signal
$s(t)$ emitted at time $t$ is received by the observer at time
$t+d(t)/{\rm c}$, and can be represented by its harmonic decomposition

\mathindent 0.2cm \begin{equation} s(t)=\sum_{n=0}^{\infty}s_{n}(t)
\hspace{0.1in} , \hspace{0.1in} s_{n}(t)=A_{n}(t)\,e^{\,2\pi jn \nu_{\rm
psr}\,\left[t+d(t)/{\rm c}\right]} \label{eq:Signal} \end{equation}

\noindent where $A_{n}(t)$ is the complex amplitude of the $n^{\rm th}$
harmonic, c the velocity of light and $j=\sqrt{-1}$. We consider binary
systems in circular orbits, for which the projected distance to the
observer is

\mathindent 0.4cm \begin{equation}
d(t)=[a_{1}\sin(i)]\cos\left(2\pi\nu_{\rm orb} t+{\rm
\phi_{orb}}\right)\,+\,{\rm Constant} \label{eq:Distance} \end{equation}

\noindent where $a_{1}$ is the pulsar's orbital radius, $i$ the orbital
inclination in radians, $\nu_{\rm orb}$ the orbital frequency and
$\phi_{\rm orb}$ the initial orbital phase measured from superior
conjunction of the pulsar. For simplicity, we shall take Constant = 0.
Expanding Eq. (\ref{eq:Distance}) in Eq. (\ref{eq:Signal}) yields

\mathindent 0.8cm \begin{equation} s_{n}(t)=A_{n}(t)\,e^{\,2\pi j
n\nu_{\rm psr} t}\,e^{\,j\varphi_n\cos\left(2\pi\nu_{\rm
orb}t+\mathrm{\phi_{orb}}\right)} \label{eq:dec} \end{equation}

\noindent where $\varphi_n = 2\pi n \nu_{\rm psr}\left[a_1\sin(i)/{\rm
c}\right]$ represents half the phase rotation in radians experienced by
the $n^{\rm th}$ harmonic of the pulsar signal during the orbit. Since
the signal $s_{n}(t)$ is the $n^{\rm th}$ harmonic of the Fourier series
decomposition, we shall study its Fourier response.

\begin{figure*} \begin{center} \begin{tabular}{@{}lr@{}}
\resizebox{8.6cm}{!}{\includegraphics{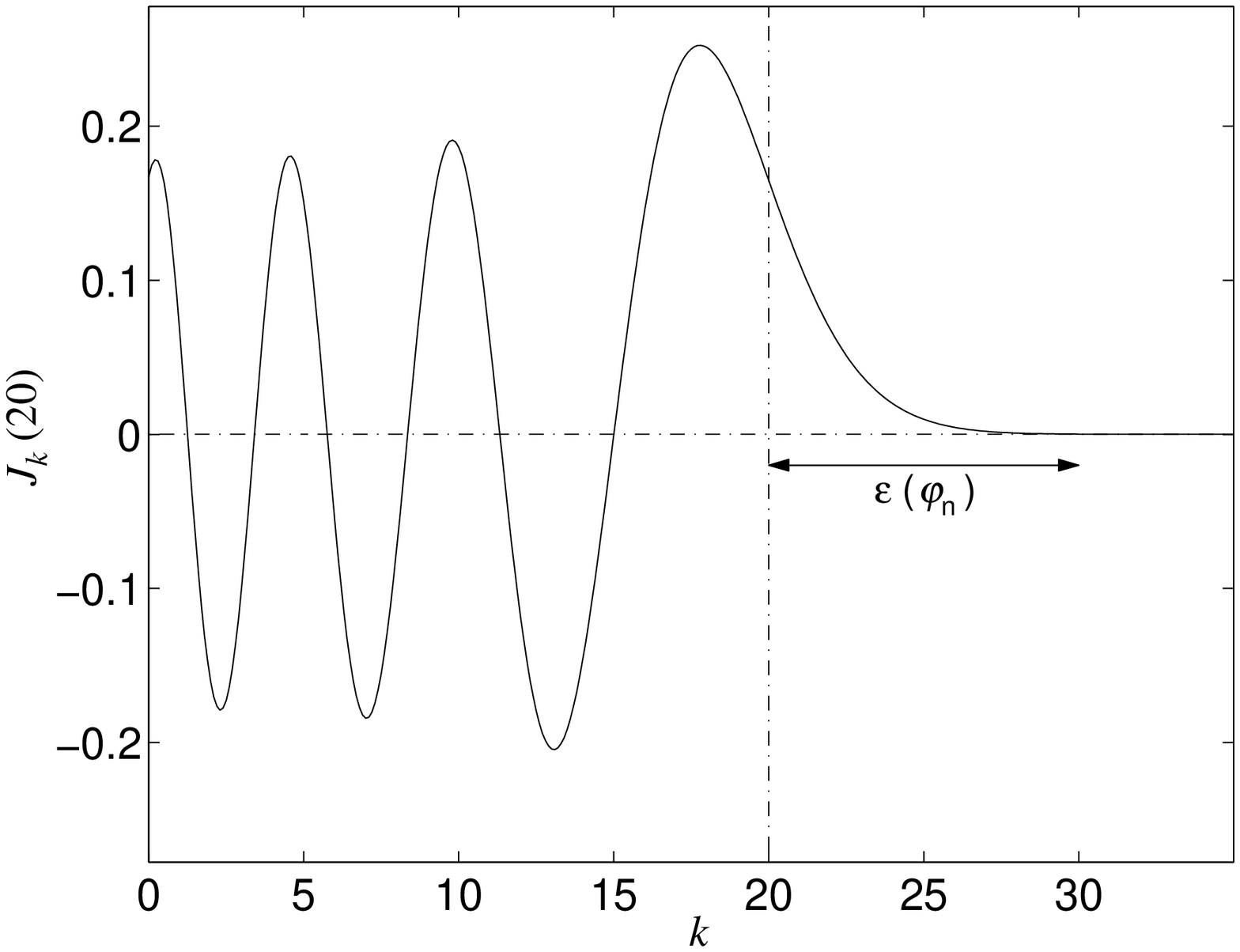}} &
\resizebox{8.6cm}{6.78cm}{\includegraphics{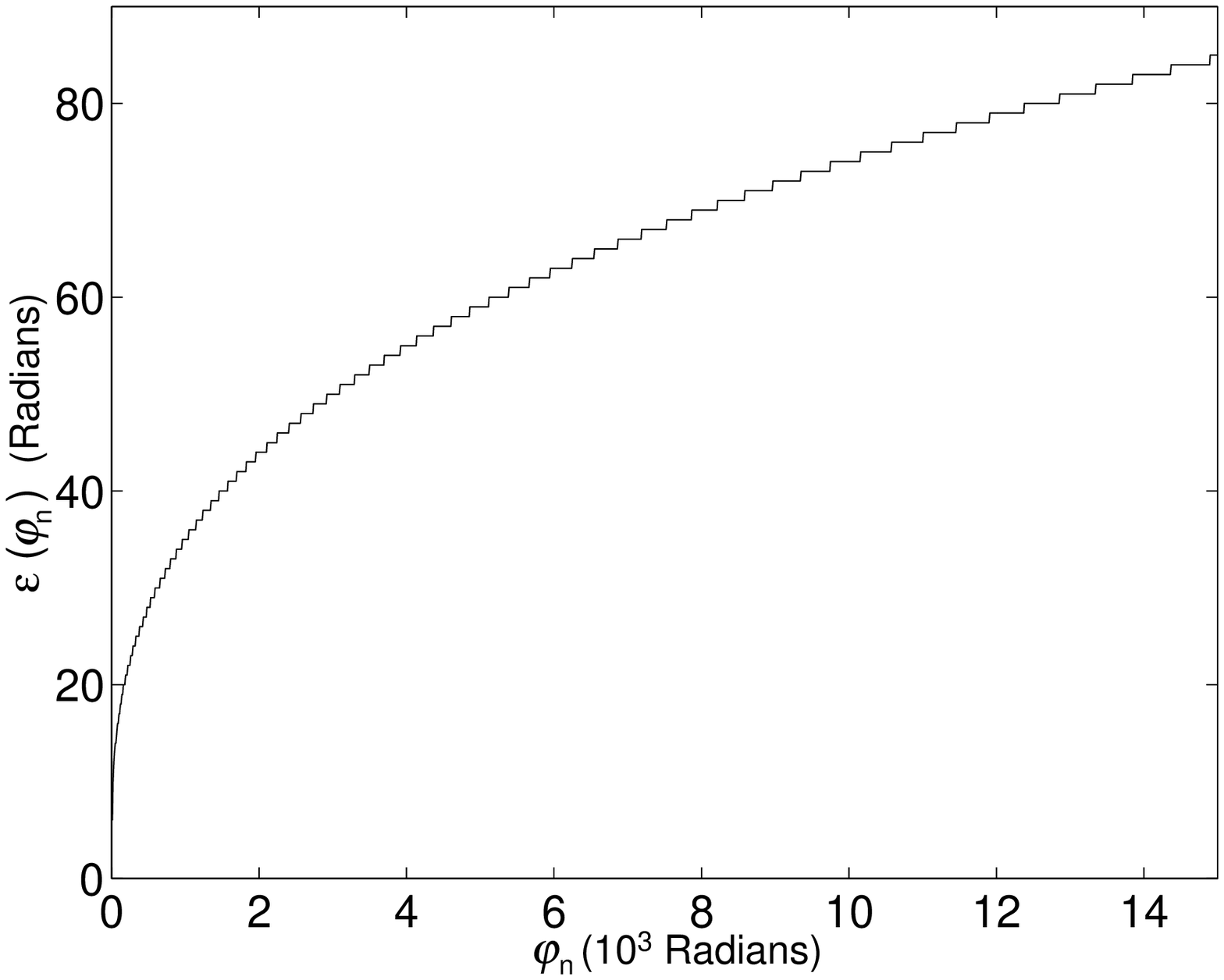}} \\
\end{tabular} \end{center} \caption[]{(a) left: $J_{k}(\varphi_n)$ with
$\varphi_{n}=20$. We plot $k \in {\Bbb R}$ whereas in reality $k \in
{\Bbb N}$. The Fourier response slightly extends beyond the amplitude of
the phase modulation as $\varepsilon(\varphi_n) \neq 0$. (b) right: in
the general case, we plot $\varepsilon(\varphi_n)$ as a function of
$\varphi_n$.} \label{fig:bessel} \end{figure*}

\section{Fourier analysis} \label{sec:fa}

We denote the Fourier response of any function $\Psi$ by $\tilde{\Psi}$
and define it by $\tilde{\Psi}(\nu)={\cal F}[\Psi(t)](\nu)$ where $\nu$
indicates the frequency and ${\cal F}$ the Fourier Transform operator.
As regards notations, Int[\,] shall define the {\it integer part of} and
Frac[\,] the {\it fractional part of}. Henceforth for consistency with
future notations, we call T-space what is usually referred to as the
time domain.

\medskip

\noindent In an ideal case, the complex amplitude $A_{n}(t) \equiv
A_{n}$, and the pulsar signal is a periodic function. However, in most
astrophysical cases the periodicity of the signal is violated because

\begin{enumerate} \item The pulsed signal strength might vary during the
observation due to eclipses, interstellar scintillation and
observational windowing. \item The pulsed signal could be modulated by
stochastic or deterministic processes such as radio pulsar nulling,
drifting subpulses or microstructure. \end{enumerate}

The latter effects are not so important because the average pulse
profile of radio pulsars is known to remain stable within the observing
window. Nevertheless in practice, $A_{n}(t)$ is a function of time but
the modulation itself remains to a large extent {\it a priori} unknown
to the observer.

\subsection{Continuous Fourier response} \label{sec:cfr}

Let $\xi_{n}(t)=e^{\,j\varphi_n\cos\left(2\pi\nu_{\rm
orb}t+\mathrm{\phi_{orb}}\right)}$ define the phase rotation factor of
Eq. (\ref{eq:dec}) and $W(t)$ a real window function of finite duration.
For an observation constrained by a window $W(t)$, Eq. (\ref{eq:dec})
can be rewritten as

\mathindent 1.6cm \begin{equation} s_{n}(t) =
A_{n}(t)\,\xi_{n}(t)\,W(t)\,e^{\,2\pi jn \nu_{\rm psr} t}\,.
\label{eq:reduced} \end{equation}

\noindent According to Fourier theory and by virtue of the shift
theorem,

\mathindent 1.1cm \begin{equation}
\tilde{s}_{n}(\nu)=\tilde{A}_{n}(\nu)*\left[
\tilde{\xi}_{n}(\nu)*\tilde{W}(\nu-n \nu_{\rm psr})\right]
\label{eq:full} \end{equation}

\noindent where $*$ denotes the convolution operator. As we aim to study
the Fourier response of the pulsating signal, we shall start with the
calculation of $\tilde{\xi}_{n}(\nu)$. By definition,

\mathindent 2cm \begin{equation}
\tilde{\xi}_{n}(\nu)=\int_{-\infty}^{\infty}\xi_{n}(t)\,e^{\,-2\pi j\nu
t}\,\mathrm{d}t\,. \label{eq:mathdef} \end{equation}

\noindent The integration of Eq. (\ref{eq:mathdef}) requires the
substitution of $\xi_{n}(t)$ by its series expansion given by (see
Appendix \ref{appendix:serie})

\mathindent 0.7cm \begin{equation} \xi_{n}(t) =
\sum_{k=-\infty}^{\infty}J_{k}(\varphi_n)\, e^{\,jk[\phi_{\rm
orb}+\pi/2]}\,e^{\,2\pi jk\nu_{\rm orb}t} \label{eq:mathserie}
\end{equation}

\noindent where $J_{k}$ are Bessel Functions of the First Kind of
integer order and $k \in {\Bbb N}$ (see also Middleditch 1981). Combining
Eqs. (\ref{eq:mathdef}) and (\ref{eq:mathserie}) yields

\mathindent 0.3cm \begin{equation}
\tilde{\xi}_{n}(\nu)=\sum_{k=-\infty}^{\infty}\left[J_{k}(\varphi_n)\,e^
{\,jk[\phi_{\rm orb}+\pi/2]}\right]\,\delta(\nu-k\,\nu_{\rm orb})
\label{eq:mathdelta} \end{equation}

\noindent where $\delta$ denotes Dirac's delta function. This
convolution greatly simplifies Eq. (\ref{eq:mathdelta}) into

\mathindent 0.8cm \begin{equation} \tilde{\xi}_{n}(\nu) = \left\{
\begin{array}{cl} J_{k}(\varphi_n)\,e^{\,jk[\phi_{\rm orb}+\pi/2]} &
\hspace{0.1in} \nu = k\,\nu_{\rm orb} \\ 0 & \hspace{0.1in}
\mathrm{otherwise} \end{array} \right.\,. \label{eq:response}
\end{equation}

\noindent In the ideal case of an infinite and continuous observation,
$\tilde{\xi}_{n}(\nu)$ represents the Fourier response of the $n^{\rm
th}$ harmonic of a signal of constant amplitude emitted by a pulsar
whose phase is modulated by circular orbital motion. Following Eq.
(\ref{eq:full}), $\tilde{\xi}_{n}(\nu)$ is defined in Eq.
(\ref{eq:response}) relative to the $n^{\rm th}$ harmonic frequency
taken as the origin of the frequency scale. Of particular interest is
its power spectrum, defined by
$P_{\xi,n}(\nu)=|\tilde{\xi}_{n}(\nu)|^{2}$. A number of important
properties of $P_{\xi,n}$ are :

\begin{enumerate}

\item Periodicity

\smallskip

$P_{\xi,n}$ is a {\it periodic sequence} with period $\nu_{\rm orb}$
whose amplitude is modulated by $J_{k}^{2}(\varphi_n)$.

\smallskip

\item Compactness

\smallskip

Since $k$ is an integer, $J_{-k}(\varphi_n)=(-1)^{k}\,J_{k}(\varphi_n)$
hence $P_{\xi,n}$ is symmetric with respect to $k=0$. Furthermore,
$\exists\, \varepsilon(\varphi_n)\;,\; \forall\,k >
[\varphi_n+\varepsilon(\varphi_n)]\;:$

\mathindent 1.3cm \[ |J_{k}(\varphi_n)| < \left|\frac{{\rm
Max}[J_{k}(\varphi_n)]_{k}}{1000}\right| \]

(see Abramowitz \& Stegun 1974 and Fig.~\ref{fig:bessel}(a)). As shown in
Fig.~\ref{fig:bessel}(b), $\varepsilon(\varphi_n) \ll \varphi_n$  so
$J_{k}(\varphi_n)$ only slightly extends beyond its argument. Therefore,
$P_{\xi,n}$ is compact on the interval $I(\varphi_n,\nu_{\rm
orb})=[-{\rm E}(\varphi_n)\,\nu_{\rm orb}\,{\rm ;}\,{\rm
E}(\varphi_n)\,\nu_{\rm orb}]$ where ${\rm
E}(\varphi_n)~=~\varphi_n~+~\varepsilon(\varphi_n)$ radians. Note that

\mathindent 2cm \begin{equation} \varphi_n\,\nu_{\rm orb} = \frac{v_{\rm
max}}{\rm c}\,n\,\nu_{\rm psr} \end{equation}

\noindent where $v_{\rm max}$ denotes the maximum radial velocity of the
pulsar. Thus, $\varphi_n\,\nu_{\rm orb}$ represents the amplitude of the
Doppler shift experienced by the $n^{\rm th}$ harmonic of the pulsar.
$P_{\xi,n}$ is a comb of $\sim 2\,\varphi_n$ sidebands with a spacing of
$\nu_{\rm orb}$. \end{enumerate}

\begin{figure}[t] \resizebox{8.6cm}{7.0cm}{\includegraphics{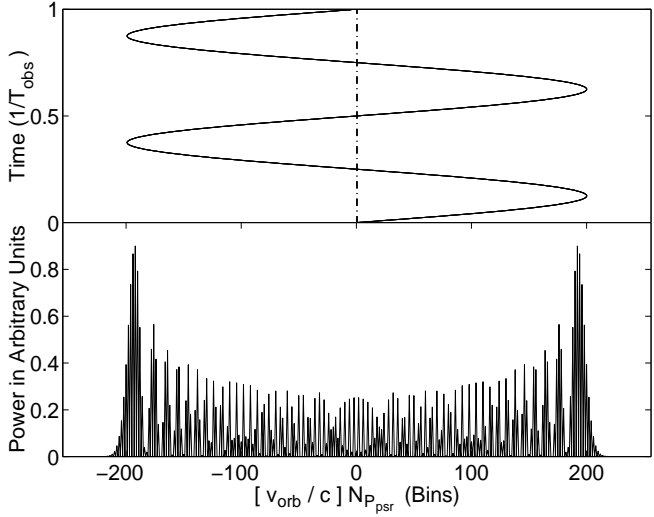}}
\caption[]{Top : pulsar radial velocity in units of Fourier frequency
bins (v$_{\rm orb}$ denotes the radial velocity of the pulsar). Bottom :
power spectrum of the 1$^{\rm st}$ harmonic shifted back to the origin
of the frequency scale ($P_{\zeta,1}$) in arbitrary units. Here,
$\varphi_{1}=100$, $N_{P_{\rm orb}}=2.0$ and $\phi_{orb}=0$. The
response is symmetric because the pulsar signal is periodic in the
observing window and $N_{P_{\rm orb}}$ is an integer.}
\label{fig:cirFig} \end{figure}

\subsection{Discrete Fourier response} \label{sec:dfr}

By necessity, any observed signal is of finite extent. The extent may be
adjustable and selectable, but must be finite. Processing a
finite-duration observation by estimating its complex spectrum directly
from the Fourier Transform encounters many difficulties which make a
binary pulsar's Fourier response more complicated.

\subsubsection{Frequency sampling} \label{sec:fs}

In practice, we define

\mathindent 1.2cm \[ \nu \equiv \nu_{\ell} \hspace{0.1in} {\rm and}
\hspace{0.1in} \nu_{\ell} =  \frac{\ell}{T_{\rm obs}} \, ,
\hspace{0.1in} \ell \in {\Bbb N}_{\,[0,N/2-1]} \]

\noindent where $T_{\rm obs}$ is the observing time, $N$ the number of
{\it equally spaced} sampling points in the data set and $\ell$ the bin
number in the Fourier spectrum.  Following Sec. \ref{sec:cfr}, we define
$\tilde{\zeta}_{n}(\ell)$ as the discrete Fourier response of the
$n^{\rm th}$ harmonic of a signal of constant amplitude emitted by a
pulsar whose phase is modulated by circular orbital motion. We also
define its power spectrum
$P_{\zeta,n}(\ell)=|\tilde{\zeta}_{n}(\ell)|^{2}$. Henceforth for
convenience, various quantities such as orbital frequency $\nu_{\rm
orb}$ will be expressed in units of Fourier bins instead of frequencies.
Thus, $\nu_{\rm orb}$ can also be referred to as $N_{P_{\rm
orb}}=\nu_{\rm orb}\,T_{\rm obs}$ in the context of a particular
observation of length $T_{\rm obs}$. The previous results then imply
that

\begin{enumerate}

\item $P_{\zeta,n}$ is a spiky $N_{P_{\rm orb}}$-periodic sequence with
a modulated amplitude. The periodicity is now expressed in units of
Fourier bins. Also note that

\mathindent 1.7cm \begin{equation} \varphi_n\,N_{P_{\rm orb}} =
\frac{v_{\rm max}}{\rm c}\,n\,N_{P_{\rm psr}} \end{equation}

\noindent where $N_{P_{\rm psr}} = \nu_{\rm psr}\,T_{\rm obs}$ is the
number of pulse periods in the observation.

\item The Fourier response $\tilde{\zeta}_{n}$ is compact on the
interval $I(\varphi_n,N_{P_{\rm orb}})$ about the $n^{\rm th}$ harmonic.
It corresponds to the Doppler shift interval of the $n^{\rm th}$
harmonic of the pulsar signal expressed in units of Fourier bins and can
be approximated by $2\,N_{P_{\rm orb}}\,\varphi_n$ (see Fig.
\ref{fig:cirFig}).

\end{enumerate}

A $N$-point Fourier Transform is elegant when the processing scheme is
cast as a spectral decomposition in an $N$-dimensional orthogonal vector
space. However, such remarkable efficiency is compromised in practice
because the number of orbital periods and pulse periods contained in an
observation are not integers.

\subsubsection{Spectral leakage} \label{sec:sl}

From the continuum of possible frequencies $\nu$, only those which
coincide with the basis of discrete frequencies $\nu_{\ell}$ will
project onto a single complex vector in the Fourier spectrum. All other
frequencies will exhibit non zero projections on the entire basis set.
This is often referred to as spectral leakage and results from
processing finite-duration signals whose frequencies are not periodic in
the observation window $W$. This leakage can strongly degrade a signal's
Fourier response.

\noindent Multiplicative weighting windows can be used to minimize such
an effect (Harris 1978). Such windows proved very useful for
multiple-tone signal detection. However, they convey other intrinsic
properties that make the detection of noisy signals difficult:

\begin{itemize} \item The Equivalent Noise Bandwidth\footnote{The
Equivalent Noise Bandwidth is the width of an ideal rectangular filter
which would accumulate the same noise power from white noise as the
window function's kernel with the same peak power gain.} increases.
\item The independent noise samples become correlated and the estimation
of the probability density function of the noise consequently increases
in complexity. \item If the windows are applied to overlapping
partitions of the data sequence, the noise samples become even more
correlated. The work load also increases. \end{itemize}

\noindent Hence we compute the Fourier response given by the straight
FFT of the data set. We can now rewrite Eq. (\ref{eq:full}) for the case
of a discrete Fourier Transform as $\tilde{s}_{n}(\ell) =
\tilde{A}_{n}(\ell) * [\tilde{\zeta}_n(\ell) * \delta(\ell-n\,N_{P_{\rm
psr}})]$ where

\mathindent 0.65cm \begin{eqnarray} \tilde{\zeta}_n(\ell) =
\sum_{k=-{\rm Int}[E(\varphi)]}^{{\rm Int}[E(\varphi)]} &
\left[\rule{0cm}{0.5cm}
J_{k}(\varphi_n)\,e^{\,jk[\phi_{orb}+\pi/2]}\,\times \right. \nonumber
\\ &  \left. \tilde{W}(\ell-k\,N_{P_{\rm orb}})
\rule{0cm}{0.5cm}\right]\,. \label{eq:snc} \end{eqnarray}

\noindent $\tilde{\zeta}_n$ can be calculated for virtually any window
$W$ provided its Fourier response is known. In that sense, Eq.
(\ref{eq:snc}) represents the analytical solution to the problem of
Fourier analysis of noiseless binary pulsar signals in the solar system
barycentric frame. In practice, $W$ is defined by a Boxcar function
whose Fourier response is given by

\mathindent 0.2cm \begin{equation} \tilde{W}_{B}(\ell) = T_{\rm
obs}\,e^{\,-j\pi\ell}\,{\rm sinc}\,(\ell) \hspace{0.1in} ,
\hspace{0.1in} {\rm sinc}\,(\ell) = \frac{\sin\,(\pi\ell)}{\pi\ell}\,.
\end{equation}

\subsubsection{Computational cost} \label{sec:ccost}

Obviously, $\tilde{\zeta}_n$ can be computed using Eq. (\ref{eq:snc}).
The sine, cosine and Bessel functions can be pre-computed and stored in
static lookup tables. The computational cost of estimating
$\tilde{\zeta}_n$ over its compact interval for the simple case of the
Boxcar function amounts to $\sim$~$50\;N_{P_{\rm orb}}E(\varphi_n)^{2}$
operations.

\noindent Alternatively, $\tilde{\zeta}_n$ can be evaluated numerically
by computing a Fast Fourier Transform (FFT) of a simulated data set of
length $K = 4$\^{}${\,{\rm Int}\left[\,\log_{2}I(\varphi_{n},N_{P_{\rm
orb}})\,\right]}$ defined by

\mathindent 0.85cm \begin{eqnarray} \zeta_{n}(k) & = & \exp\left[2\pi
jn\left(\frac{K}{2}+\eta \right)\frac{k}{K}+ \right. \nonumber \\ & &
\left. j\varphi_{n}\cos\left(2\pi N_{P_{\rm orb}}\frac{k}{K}+\phi_{\rm
orb}\right)\right] \label{eq:comp} \end{eqnarray}

\noindent where $k \in [0\,{\rm ;}\,K-1]$. The spectral leakage factor
is defined by $\eta={\rm Min}[{\rm Frac}[\nu_{\rm psr}T_{\rm
obs}],1-{\rm Frac}[\nu_{\rm psr}T_{\rm obs}]]$ such that $\eta \in
[-0.5\,{\rm ;}\,0.5]$.  The number of operations required in that case
to estimate $\tilde{\zeta}_n$ reduces to $\sim 2\,K(5+\log_{2}K)$. We
clearly favor the latter solution as the computational cost is much
smaller.

\subsubsection{Additive noise contribution} \label{sec:anc}

Let us now consider the effect of including additive white noise with
the pulsar signal. The variance of the noise is assumed equal at all
frequencies. Since the noise is additive and the Fourier Transform
operator is linear, the noisy power spectrum is given by

\mathindent 1.8cm \begin{equation} {\cal P}_{n}(\ell) =
P_{\zeta,n}(\ell) + |\tilde{x}(\ell)|^{2} + \Lambda(\ell) \end{equation}

\noindent where $\tilde{x}(\ell)$ represents the Fourier decomposition
of the noise samples at the frequency corresponding to bin $\ell$.
$\Lambda(\ell)$ denotes the cross-terms between the signal and noise
complex amplitudes :

\mathindent 1.4cm \begin{equation} \Lambda =
2\,\left[\Re(\tilde{s}_{n})\times\Re(\tilde{x})+\Im(\tilde{s}_{n})\times
\Im(\tilde{x})\right]\,. \end{equation}

\noindent For a wide range of probability distributions, real ($\Re$)
and imaginary ($\Im$) parts of the Fourier decomposition of the noise
samples are gaussian distributed (Jenkins \& Watts 1968), hence
$|\tilde{x}(\ell)|^{2}$ will, with proper normalization, be chi-squared
distributed with 2 degrees of freedom. The proper normalization factor
$\beta$ for a Discrete Fourier Transform (DFT) of length $K$ defined by

\mathindent 2cm \begin{equation} \tilde{\Psi}(\ell) \equiv
\frac{1}{\beta}\sum_{k=0}^{K-1}\;\Psi(k)\,e^{\;-2\pi j\ell k/K}
\end{equation}

\noindent depends on the initial noise probability density function.
This function is gaussian in radio observations and poissonian in X-ray
observations hence we find $\beta~=~\sqrt{K/2}$. When the noise is
chi-squared distributed, we find $\beta~=~\sqrt{2K}$.

\noindent Because of the orthogonality of the Fourier decomposition,
$\Re$ and $\Im$ are independent. Therefore, $\Lambda(\ell)$ is gaussian
distributed with mean $\mu(\ell) = 0$ and $\sigma(\ell) =
\sqrt{2\;P_{\zeta,n}(\ell)}$. The {\it statistical expectation} of
$\Lambda$ consequently reduces to zero.

\begin{figure} \resizebox{8.6cm}{!}{\includegraphics{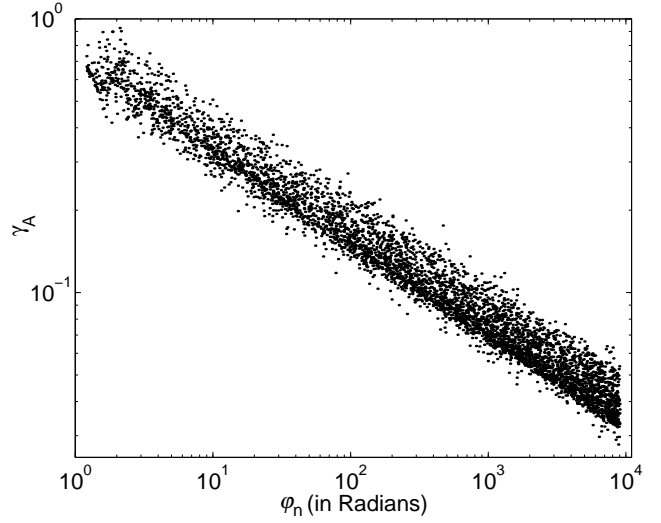}}
\caption[]{Efficiency factor $\gamma_{\rm A}$ obtained from numerical
simulations. The plot shows 5000 binary pulsar systems with random
parameters. The distributions are uniform in the ranges given by
$N_{P_{\rm orb}} \in [2.0;20.0]$, $\varphi_{n} \in [1.2;9000]$, $\eta
\in [-0.5;0.5]$ and $\phi_{orb} \in [-\pi;\pi]$.} \label{fig:binaryLoss}
\end{figure} 

\subsection{Summary}

We have now completed our Fourier analysis of a periodic signal whose
phase is modulated by circular orbital motion. For an observation of
finite duration, we can summarize our results as follows~:

\begin{itemize}

\item The basic analytical solution is given by Eq. (\ref{eq:snc}).

\item The power spectrum is $N_{P_{\rm orb}}$-periodic over a finite
frequency range about each harmonic.

\item The phase in Fourier space is unpredictable {\it a priori} because
$N_{P_{\rm orb}}$ and $\phi_{\rm orb}$ are unknown.

\item A numerical simulation including an FFT is more efficient than a
direct computation of Eq. (\ref{eq:snc}).

\item The statistical properties of white noise do not favor any
particular periodicity. The particular signature of a pulsar signal
whose phase is modulated by a circular orbital motion consequently
provides an opportunity for an incoherent detection of weak pulsar
signals in broadband noise.

\end{itemize}

\begin{figure*} \begin{center} \begin{tabular}{@{}lr@{}}
\resizebox{8.6cm}{!}{\includegraphics{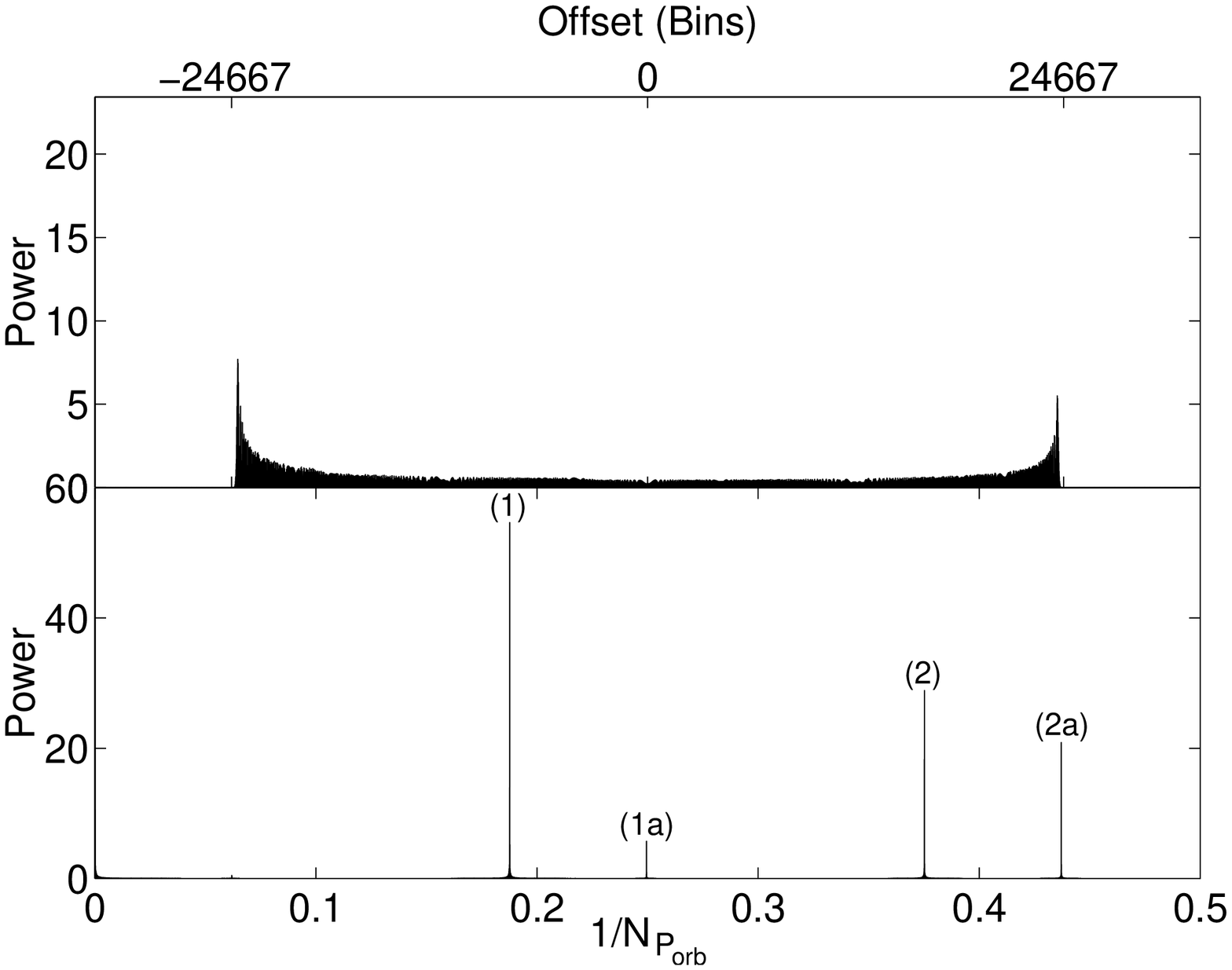}} &
\resizebox{8.6cm}{!}{\includegraphics{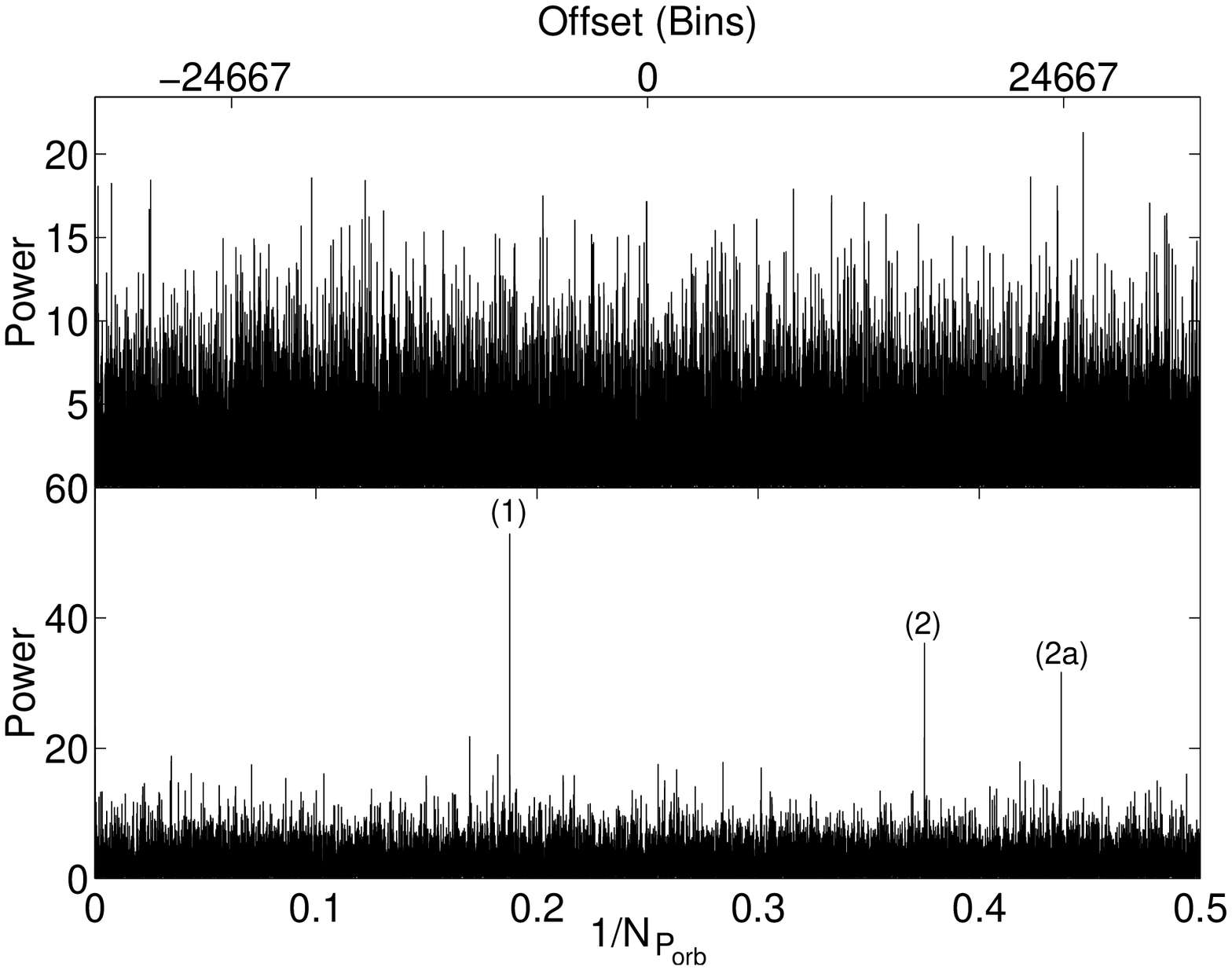}} \\ \end{tabular}
\end{center} \caption[]{The binary system is described by $\nu_{\rm psr}
= 1$ KHz, $P_{\rm orb} = 1.5$ hours, M$_2$ = 0.8 M$_{\odot}$ and
$\phi_{orb} = 53^o$ so that $\varphi_1 = 4571$ radians, $N_{P_{\rm orb}}
= 5.34$ and $\eta = 0.23$. Upper plots show raw power spectra as
obtained from a direct DFT of a noiseless/noisy observation (left/right
respectively). Lower plots show ${\cal Q}$. The pulsar is detected at
$1/N_{P_{\rm orb}}$ as shown by feature (1). The second harmonic is
indicated by feature (2). Features (1a) and (2a) are essentially caused
by beating of $N_{P_{\rm orb}}$ with bins in ${\cal P}$-space. The
relative position of (1a)/(1) and (2a)/(2) is given by 1+Frac$[N_{P_{\rm
orb}}]$. Multiplicative weighting windows could almost remove such
features caused by spectral leakage effects (see Sec. \ref{sec:sl}).}
\label{fig:exsn} \end{figure*}

\noindent We have considered pulsars in circular orbits. The case of
eccentric orbits is extremely difficult to solve analytically. The
envelope of the  Fourier response becomes very complicated and can not
be described by simple Bessel functions. However, the sideband spacing
still remains the same and therefore produces a periodic sequence as in
the case of circular orbits.

\noindent We have not considered non-uniform sampling as may happen for
long observations due to Doppler correction for the observatory, either
earthbound or satellite-borne.  We assumed that the time series has been
resampled with equal intervals for an observation in the solar system
barycentric frame.

\noindent In Sec. \ref{sec:pst}, we briefly present three pulsar search
techniques widely used in radio/X-ray astronomy and investigate their
limiting sensitivity to binary pulsars in close circular orbits.  In
Sec. \ref{sec:pcrt}, we use the results of the present section to
develop a very efficient implementation of the Phase Modulation
Searching method (Ransom 1999).  Eventually, we compare the efficiency of
these recovery techniques and discuss the detectability of a range of
binary pulsars in ultra-short circular orbits.

\section{Searching for pulsars in close binary systems}
\label{sec:pst}

When a pulsar signal is present in a time series, which in the case of
radio observations has already been corrected for the dispersion due to
its passage through the interstellar medium, it manifests itself as a
set of discrete harmonics in the Fourier spectrum. The number of
significant harmonics depends on the duty cycle of the pulsar
(fractional on-time of the pulsar signal). Whereas ordinary pulsars have
a typical duty cycle of about 4\% (indicating about 12 significant
harmonics in the spectrum), the millisecond pulsars seem to have a duty
cycle of about 10\% (about 5 harmonics). In the population of X-ray
pulsars, the High Mass X-ray Binary pulsars have complex pulse profiles
and thus have a few significant harmonics. However, some Low Mass X-ray
Binary pulsars, including the only known millisecond pulsar among them,
show nearly sinusoidal pulsations. Naturally, whenever we have more than
one harmonic, ``harmonic-folding'' (Bhattacharya 1998) will improve the
probability of detection when compared to looking for statistically
significant isolated harmonics in the power spectrum.

\noindent In the case of solitary pulsar signals, harmonics are
essentially confined to one bin whose power is decreased by ${\rm
sinc}^{2}({\rm Min}[|{\rm Frac}[n\eta]|,1-|{\rm Frac}[n\eta]|])$, where
$n$ is the harmonic number\,; it is important to account for the
spectral leakage factor $\eta$ because we will compare different
recovery techniques that are not equally affected by such an effect. In
a way similar to that of Johnston \& Kulkarni (1991), we define
the efficiency factor $\gamma$, such that a binary pulsar must, on
average, carry $\gamma^{-1}$ more pulsed flux density than its virtual
solitary counterpart to be detected with the same significance level. We
now present several methods and calculate their $\gamma$ either from
theoretical predictions or numerical simulations. We will compare and
discuss their performances in fuller details in Sec. \ref{sec:dis}.

\bigskip

\begin{itemize}

\item {\bf Method A}

\medskip

Consider a signal emitted by a pulsar in a binary system with a circular
orbit. This signal is corrected to the solar system barycenter if
necessary and uniformly resampled. Then, we simply search for
statistically significant peaks in the discrete power spectrum --
henceforth referred to as ${\cal P}$-space. As seen in Sec.
\ref{sec:fs}, a given harmonic number $n$ will be spread out over
$I(\varphi_n,N_{P_{\rm orb}})$ bins. The efficiency factor $\gamma_{\rm
A}$ is plotted as a function of $\varphi_{n}$ for 5000 systems with
random parameters in Fig. \ref{fig:binaryLoss}. The distributions are
uniform in the ranges given by $N_{P_{\rm orb}} \in [2.0;20.0]$,
$\varphi_{n} \in [1.2;9000]$, $\eta \in [-0.5;0.5]$ and $\phi_{orb} \in
[-\pi;\pi]$. Clearly, $\gamma_{\rm A}$ is strongly correlated with
$\varphi_{n}$. The dispersion about the least-square-fit line

\mathindent 1.8cm \begin{equation} \gamma_{\rm A} \simeq [0.74\pm
0.1]\,\varphi_{n}^{-0.325} \label{eq:caseag} \end{equation}

\noindent is mostly caused by spectral leakage effects. Therefore,
$\gamma_{\rm A}$ is maximized when $N_{P_{\rm orb}}$ and $\eta$ are
close to integer values.

\noindent As $\varphi_{n} \propto a_{1}$ and $\gamma_{\rm A} $ varies
little with orbital period, we can see in Fig. \ref{fig:binaryLoss} that
close binaries are easier to detect than wide binaries. This is because
a pulsar with a relatively small $\varphi_n$ experiences less
acceleration.

\noindent In the worst case, sensitivity limits are decreased by a
factor of about 30 when compared to a solitary pulsar.

\medskip

\item {\bf Method B }

\medskip

Numerous authors have developed techniques to correct for the Doppler
smearing caused by the orbital motion. However, a complete coherent
recovery involves searching in a parameter space defined by a large
range of keplerian parameters. This is computationally unfeasible for
the present day technology. As a compromise, some authors (Middleditch
\& Priedhorsky 1986; Anderson et al. 1990; Wood et al. 1991; Camilo
et~al. 2000; Johnston \& Kulkarni 1991) have attempted a constant
acceleration correction. Our simulations showed, in good agreement
with Camilo et\,al. (2000), that almost 100\% of the pulse strength is
on average recoverable for all orbital phases if $N_{P_{\rm orb}}
\lsim 1/7$. The efficiency of constant acceleration searches can
therefore be expressed as

\mathindent 2.15cm \begin{equation} \gamma_{\rm B} \simeq
\frac{1}{\sqrt{7\;N_{\rm P_{\rm orb}}}} \label{eq:linacg}\,.
\end{equation}

\noindent Clearly, constant acceleration searches perform well when the
observing time is much less than the orbital period. The efficiency of
such a method can be expressed in various ways. We have estimated
$\gamma_{\rm B}$ in order to maximize the recovery for the smearing due
to the time-dependent Doppler effect experienced by the binary pulsar.
This method basically allows for searching the whole time series with
negative and positive acceleration values, including zero acceleration.
Therefore in practice, it includes Method A which may itself prove more
sensitive when $N_{\rm P_{\rm orb}}$ is large. However, Method B and
$\gamma_{\rm B}$ will henceforth refer to Eq. (\ref{eq:linacg}) as
aiming for a ``non zero'' constant acceleration correction.

\medskip

\item {\bf Method C}

\medskip

Another method widely used in X-ray astronomy consists of dividing a
long time series into $N_{\rm s}$ sub-segments in order to reduce the
frequency resolution so that the Doppler shifted pulsar signal remains
essentially confined to one bin in ${\cal P}$-space (van der Klis
1989). The $N_{\rm s}$ short power spectra are then incoherently
summed. The stacked power spectrum is subsequently searched for
statistically significant peaks.

\noindent Using confidence levels instead of signal-to-noise ratio
(SNR) is of great importance here because the noise distribution
itself is modified in this process. The efficiency factor $\gamma_{\rm
C}$ can therefore be measured as follows : let us define
$\chi_{\alpha}^{D_f}$ as the inverse cumulative density function for a
chi-squared distribution with $D_f$ degrees of freedom at the
confidence value $\alpha$ ($D_f = 2\,N_{\rm s}$). By analogy with
Sec. \ref{sec:anc}, we can define $X=\sum_{k=1}^{N_{\rm
s}}|\tilde{x}_k|^{2}$ so that
Prob$(X>\chi_{\alpha}^{D_f})=\alpha$. The average expectation of $X$
is given by $\chi_{1/2}^{D_f}$. When $D_f$ is large, the noise
distribution approaches a gaussian and $\chi_{1/2}^{D_f}$ coincides
with the average value of $X$. Also, when $D_f$ is large, the noise
can be considered additive in power instead of in Fourier amplitude
(Vaughan et al. 1994; Groth 1975).  Hence, $\gamma_{\rm C}$ is given
by

\mathindent 1.5cm \begin{equation} \gamma_{\rm C} = \sqrt{
\begin{array}{lcl} \chi_{\alpha}^{2} & - & \chi_{1/2}^{2} \\ [0.15cm]
\hline \\ [-0.25cm] \chi_{\alpha\,N_{\rm s}}^{2\,N_{\rm s}} & - &
\chi_{1/2}^{2\,N_{\rm s}} \end{array} } \label{eq:gad}\,. \end{equation}

$\gamma_{\rm C}$ must be evaluated at discrete values given by $N_{\rm
s} = 2^p, p \in {\Bbb N}$. In a sense, such an approach is very simple
as $\gamma_{\rm C}$ depends only on $I(\varphi_n,N_{P_{\rm orb}})$, but
we can not immediately recover any orbital parameter. However, this
method can be very powerful (see Sec. \ref{sec:dis} for a fuller
discussion).

\end{itemize}

\section{Partial coherence \\ recovery technique} \label{sec:pcrt}

We now describe our Partial Coherence Recovery Technique. We start with
an observation of radio/X-ray pulsations which has been Fourier
Transformed as in Method A. Somewhere in ${\cal P}$-space lies a set of
$N_{P_{\rm orb}}$-periodic sequences centered on harmonic frequencies of
$N_{P_{\rm psr}}$. This periodicity, often buried beneath the noise, can
be detected by computing a short DFT of ${\cal P}_n$ of length $K \sim
I(\varphi_n,N_{P_{\rm orb}})$ bins. Subsequently, we define ${\cal
Q}=|\tilde{{\cal P}}_n|^2$, hereby defining ${\cal Q}$-space as some
sort of 'incomplete' T-space (T-space is obtained by an inverse DFT of
Fourier {\it amplitudes} whereas we perform a forward DFT of Fourier
{\it powers}). Searching for statistically significant peaks in ${\cal
Q}$ would reveal the presence of a sufficiently strong pulsar signal at
$1/N_{P_{\rm orb}}$ if $N_{P_{\rm orb}} > 2$ (see Shannon's Sampling
Theorem - Shannon 1948). Thus, a detection is made as shown in Fig.
\ref{fig:exsn}. Both pulsar and orbital phases are lost in ${\cal
Q}$-space, hence we call this technique Partial Coherence Recovery
Technique (henceforth referred to as PCRT and Method D).

\noindent The following sections are devoted to pulsar signal detection
and estimation, where detection is the task of determining if a pulsar
signal is present in an observation, while estimation is the task of
obtaining the values of the parameters describing the signal.

\subsection{Detection} \label{sec:detect}

Since ${\cal P}_n$ is localized in ${\cal P}$-space and its extent
unknown {\it a priori}, we must compute many DFTs at all possible
harmonics of $N_P{_{\rm psr}}$ with various widths $K$. We subsequently
use a set of sliding windows ${\cal W}$ and define ${\cal W}_{K,\ell}$
as the window of size $K$ centered on bin $\ell$. A sufficiently strong
pulsar signal will be detected when ${\cal W}_{K,\ell}$ contains ${\cal
P}_n$, i.e. $\ell \approx N_{P_{\rm psr}}$ and $K \gsim
I(\varphi_n,N_{P_{\rm orb}})$.

\begin{table} \centering \caption{Fourier spectrum statistics. The
kurtosis is measured on real and imaginary Fourier samples obtained from
an exponentially distributed noise. A normal distribution has Kurtosis =
3. Higher values indicate leptokurtic distributions.}
\begin{tabular}{@{}cccccccc@{}} $K$ & 16 & 32 & 64 & 128 & 256 & 512 &
1024 \\ \hline Kurtosis & 3.83 & 3.51 & 3.25 & 3.10 & 3.05 & 3.02 & 3.01
\\ \end{tabular} \label{tab:kur} \end{table}

\subsubsection{Sizes of the short DFTs} \label{sec:sfft}

Since $P_{\zeta,n}$ is compact, $K$ must be $\sim
I(\varphi_{n},N_{P_{\rm orb}})$ bins to achieve optimal detectability.
Computing DFTs of random sizes is not achieved efficiently with modern
algorithms, so we must compromise our sensitivity to minimize the
computational effort by computing the DFTs using FFTs of dyadic length
$K = 2^p, p \in {\Bbb N}$.

\noindent We need a well-behaved noise probability density function to
compute appropriate detection levels. The noise distribution in ${\cal
Q}$-space will approach gaussian statistics when $K$ is large enough
(Central Limit Theorem). In Table \ref{tab:kur}, we show that such a
distribution is leptokurtic for small values of $K$. This implies that
more probability is distributed in the tail, so a larger number of high
noise values must be expected. In other words, a search code would pick
up more noise samples than we would expect on the basis of a gaussian
distribution. The lower limit to assign to $K$ is however somewhat
arbitrary as one has to define a criterion of exclusion where an
estimator deviates from gaussian statistics; we chose $K_{\rm min} =
2^8$.

\noindent The upper limit is determined by ${\rm Max}
[I(\varphi_{n},N_{P_{\rm orb}})]$. It corresponds to ``the'' extreme
astrophysical case where the number of bins over which the signal power
is distributed is maximum. Let us consider a hypothetical 0.5 ms pulsar
of 1.4 M$_{\odot}$ with a companion star of 1.4 M$_{\odot}$ in a binary
system with a circular orbit of 3.728 hours; $i = 90$ degrees,
$N_{P_{\rm orb}}$ = 2 and $n$ = 2. We use the second harmonic as we
consider an extreme case and therefore aim to maximize $\varphi_n$. The
orbital velocity of the pulsar $v_{\rm max} = 279$ km/s and the second
harmonic power spreads over about $200\times 10^{3}$ ($\lsim 2^{18}$)
bins. We therefore take an upper limit $K_{\rm max}$ of $2^{18}$.

\noindent The number of values possibly taken by $K$ is given by $N_{K}
= 1 + \log_2 K_{\rm max} - \log_2 K_{\rm min}$. In the case we
considered above, $N_{K} = 11$.

\begin{figure}[t] \resizebox{8.6cm}{!}{\includegraphics{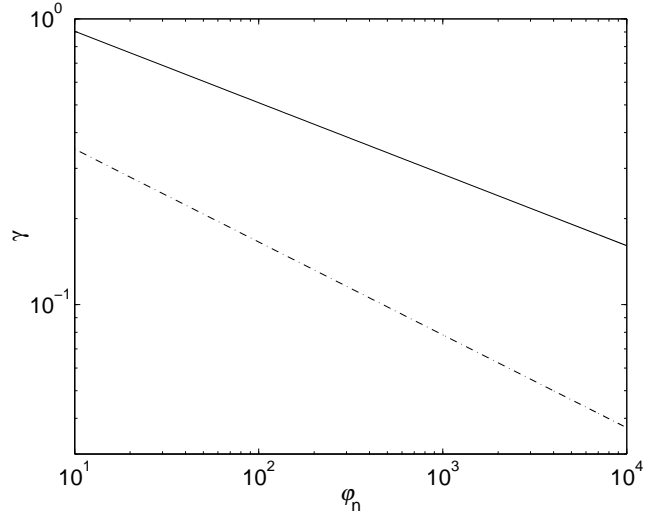}}
\caption[]{The dot-dashed curve (Method A) is given by Eq.
(\ref{eq:caseag}). The solid curve represents Method D (PCRT).}
\label{fig:sens1} \end{figure}

\subsubsection{Sliding windows} \label{sec:sw}

The ideal situation occurs when the power spectrum of ${\cal
W}_{K,\ell}$ can be computed at all possible frequency bins $\ell$ where
a pulsar signal is possibly expected. Let us denote that number by $L$.
If the sampling time is 0.1 ms and minimum pulsar period 0.5 ms, $L
\approx N/5$. The number of operations required to compute all FFTs is
given by (see Appendix \ref{appendix:offt1})

\mathindent 0.9cm \begin{equation} {\cal O}_{\rm FFT} \simeq 2^{N_{K}} L
K_{\rm min}\left[\,\log_{2}K_{\rm min}+N_{K}-2\,\right]\,.
\label{eq:offt1} \end{equation}

\noindent In our example where $N_{K}=11$ and $K_{\rm min}=256$, ${\cal
O}_{\rm FFT} \approx 9\,L \times 10^{6}$. If $N = 2^{28}$, ${\cal
O}_{\rm FFT} \approx 4.8 \times 10^{14}$ which requires about 5.6 days
of computations on a 1 GigaFLOPS\footnote{$10^{9}$ floating-point
operations per second (FLOPS).} machine. This is prohibitive. An
alternative is to decrease the fraction of overlap $\varrho$ between
consecutive ${\cal W}_{K,\ell}$ where $\varrho = 1-\rho/K$, $\rho \in
[1:K_{\rm min}-1]$ (note that $\varrho$ is a fraction, whereas $\rho$
denotes a number of bins). The number of operations then reduces to (see
Appendix \ref{appendix:offt2})

\begin{figure}[t] \resizebox{8.6cm}{!}{\includegraphics{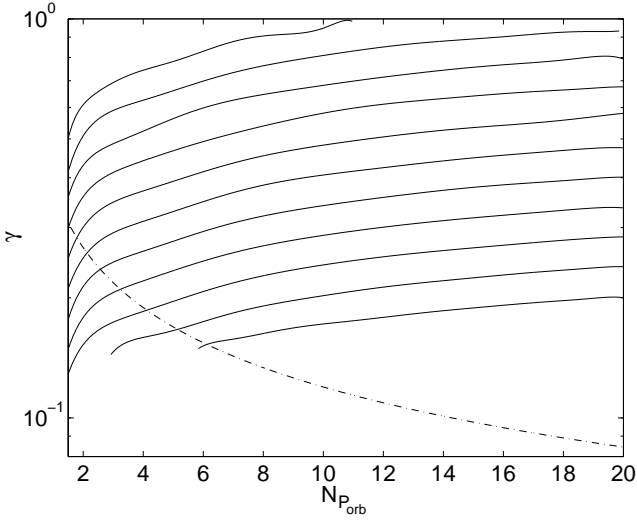}}
\caption[]{The dot-dashed curve (Method B) is given by Eq.
(\ref{eq:linacg}). Note that $\gamma_{\rm B}$ scales with $N_{P_{\rm
orb}}$. Solid curves (Method D) correspond to $K = 2^8 \ldots 2^{18}$
(upper curve is $K = 2^8$). Higher $\gamma$ implies better sensitivity.
Method D is superior to method B in nearly all cases.} \label{fig:sens2}
\end{figure}

\mathindent 0.5cm \begin{equation} {\cal O}_{\rm FFT} \simeq
\frac{L}{1-\varrho}\left[N_{K}\log_{2} K_{\rm
min}+\frac{N_{K}(N_{K}-1)}{2}\right]\,. \label{eq:offt2} \end{equation}

\noindent If $\varrho = 0.87$, ${\cal O}_{FFT} \approx 5.9 \times
10^{10}$ which requires about 1 minute of computations on a 1 GigaFLOPS
machine.

\medskip

\noindent A sliding window of length $K$ must catch $P_{\zeta,n}$ with
$I(\varphi_n,N_{P_{\rm orb}}) \leq \varrho\,K$. For such signals where
the probability of being entirely swept up by ${\cal W}_{K,\ell}$ is 1,
at least 1-$\varrho$ fraction of the sliding window will convey only
noise. In that sense and because ${\cal W}_{K,\ell}$ is evaluated for a
small number of sizes $N_{K}$, we can say that our algorithm suffers
from our limited computation power.

\noindent The number of correlated spectral samples eventually obtained
after processing $L$ bins with a single sliding window of length $K$
amounts to $L/2(1-\varrho)$. Overlapping windows cause another intrinsic
problem, namely the correlation of random components. If ${\cal
W}_{K,\ell}$ is defined with a Boxcar function, successive $\tilde{{\cal
W}}_{K,\ell}$ will be $100~\times~\varrho$~\% correlated. This implies
that the number of uncorrelated spectral samples obtained after
processing $L$ bins with a single window of size $K$ drops back to
$L/2$.

\subsubsection{Detection levels} \label{sec:dl}

${\cal W}_{K,\ell}$ contains a chi-squared distributed broadband noise,
and may also include a $N_{P_{\rm orb}}$-periodic sequence caused by a
signal. As seen in Sec. \ref{sec:dfr}, such a sequence is very spiky.
This comb of sidebands essentially decomposes into $H={\rm
Int}[N_{P_{\rm orb}}/2]$ harmonics in ${\cal Q}$-space if $N_{P_{\rm
orb}} > 2$ (Shannon 1948). The presence of a sufficiently strong binary
pulsar signal would be detected in ${\cal Q}$ at harmonics of its
orbital period as shown in Fig. \ref{fig:exsn}.

\begin{table} \centering \caption{Harmonic folding ambiguity. $L/N_H$ is
given by Eq. (\ref{eq:nh}) and $N=2^{28}$. $N_A$ denotes a number of
combinations.} \begin{tabular}{@{}lrrrrrrrrr@{}} $H$ & 1 & 2 & 3 & 4 & 5
& 6 & 7 & 8 & 9 \\ \hline $L/N_H$ & 4 & 12 & 24 & 40 & 60 & 84 & 112 &
144 & 180 \\ $N_A$ & 1 & 3 & 5 & 9 & 13 & 17 & 23 & 31 & 37 \\
\end{tabular} \label{tab:harm} \end{table}

\noindent When $N_{P_{\rm orb}} > 4$, it becomes useful to fold
successive harmonics in ${\cal Q}$ to improve detection limits. Harmonic
folding (Bhattacharya 1998) denotes the process of incoherently gathering
harmonically related signal peaks in a power spectrum. For instance in
Fig. \ref{fig:exsn}, we clearly see that adding the two harmonics,
namely features (1) and (2), will greatly improve the significance level
of the detection.

\noindent Folding means incoherent addition of $H$ independent
chi-squared distributed samples resulting in chi-squared distributed
noise with $2H$ degrees of freedom. Since a detection has a statistical
significance for a particular noise distribution, the number of
uncorrelated spectral samples $N_{H}$ to be searched for a pulsar signal
must be dealt with separately as a function of $H$ . In other words, the
significance of a power obtained after folding 2 or 3 harmonics is
different because the noise distribution itself depends on the harmonic
fold number. Therefore, we define $N_H$ as the number of samples that
can be folded a maximum of $H$ times in ${\cal Q}$-space ($H = 1$ means
no folding since we consider only harmonic number 1). A simple dichotomy
gives

\mathindent 1cm \begin{equation} N_{H} =
\frac{L}{2}\left[\frac{1}{H}-\frac{1}{H+1}\right] =
\frac{L}{2\,H\,(H+1)} \label{eq:nh} \end{equation}

\noindent where, for a given sliding window of size $K$, $L/2$
represents the total number of uncorrelated powers in ${\cal Q}$-space
and the term within square brackets denotes the fractionnal amount of
samples that can be folded a maximum of $H$ times.

\noindent Another complication arises because each frequency channel has
a finite bandwidth. We can incoherently add bin number 10 with bin
number 20, but also with bin number 19 or 21 depending whether $\eta <
-0.25$, $|\eta|<0.25$ or $\eta>0.25$. This ambiguity becomes extremely
complicated when $H$ increases and requires a large amount of
book-keeping. We define $N_A$ as a measure of such an ambiguity. In
other words, $N_A$ represents the number of possible combinations in
folding $H$ harmonics. As shown in Table \ref{tab:harm}, $N_A$ increases
with harmonic number $H$. A comprehensive search must cover all
possibilities. Therefore, the significance level of a detection obtained
after folding must take into account the number of combinations $N_A$ as
well as the appropriate noise distribution.

\noindent The detection level at harmonic number $H$ is consequently
given by $\chi_{\alpha/N_A N_H H_{\rm max}}^{2H}$, where $H_{\rm max}=9$
in our case. We used a confidence level $\alpha$ of 1\% to estimate
$\gamma_{\rm D}$.

\subsection{Estimation}

Once a significant peak is detected and localized both in ${\cal
P}$-space and ${\cal Q}$-space, we can perform efficient complex
cross-correlations of length $\sim I(\varphi_n,N_{P_{\rm orb}})$ with a
relatively small number of functions $\tilde{\zeta}_n$ given by Eq.
(\ref{eq:comp}). Such a coherent recovery of a suspected pulsar signal
is very computationally efficient and extremely sensitive. As we showed
in Sec. \ref{sec:ccost}, $\zeta_n$  is defined by four parameters :

\begin{enumerate}

\item $P_{\rm orb}$ is obtained directly from ${\cal Q}$ where a pulsar
signal has first been suspected as in Fig. \ref{fig:exsn}. The accuracy
to which this parameter is known improves with $\varphi_n$ and $T_{\rm
obs}$.

\item $\varphi_{n}$ can be approximately estimated given $N_{P_{\rm
orb}}$ and the length $K$ of the current sliding window. However, a more
accurate estimation can be obtained efficiently by performing sliding
FFTs of various lengths about $K$ in order to maximize the SNR at the
frequency corresponding to $1/N_{P_{\rm orb}}$. Since we assumed
circular orbits $P_{\xi,n}$ is symmetric, so this process also yields an
approximate value of $\nu_{\rm psr}$.

\item $\phi_{\rm orb}$ can be obtained by performing complex
cross-correlations in T-space. Since we work on a small number of
Fourier amplitudes (as $I(\varphi_n,N_{P_{\rm orb}}) \sim 2\,N_{P_{\rm
orb}}\,\varphi_n$ bins), we can inverse Fourier Transform ${\cal Q}$ and
complex cross-correlate with our reference function $\zeta_n$ given by
Eq. (\ref{eq:comp}) and $\phi_{\rm orb}=0$. The initial orbital phase
$\phi_{\rm orb}$ can be easily recovered if $N_{P_{\rm orb}}$ is known.
In practice $\nu_{\rm psr}$ is approximated when computing $\zeta_n$.

\item $\eta$ gives $\nu_{\rm psr}$ within about half a frequency bin of
precision. This parameter can be obtained by complex cross-correlating
${\cal Q}$ with $\tilde{\zeta}_n$ using previous estimates of $N_{P_{\rm
orb}}$, $\varphi_{n}$ and $\phi_{\rm orb}$. \end{enumerate}

\noindent The pulsar frequency and all measurable binary parameters are
therefore known at the end of this procedure. Such a recipe holds for
pulsars in circular orbits. If the periodic sequence in ${\cal P}$-space
is caused by a pulsar in eccentric orbit, another procedure must be used
and further processing is required. We have not investigated such a case
as our aim is to give a complete recipe for searching for pulsars with
circular orbital motion. The case of eccentric orbits should however
receive full attention in the context of the Phase Modulation Searching
method.

\section{Discussion} \label{sec:dis}

\begin{figure}[t!]
\resizebox{8.6cm}{!}{\includegraphics{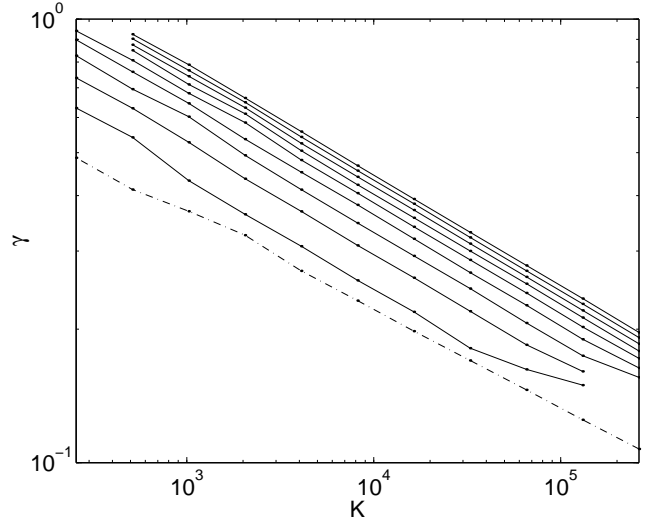}} \caption[]{The
dot-dashed curve (Method C) is given by Eq. (\ref{eq:gad}). $\gamma_{\rm
C}$ depends only on $K \sim 2\,N_{P_{\rm orb}}\,\varphi_n$. Solid curves
(Method D) correspond to $N_{P_{\rm orb}} \in [2\,H;2\,(H+1)[$ where $H
= 1 \ldots 9$ is the harmonic fold number defined in Sec. \ref{sec:dl}.
The upper curve corresponds to $H=9$. The $\gamma$ values are estimated
at $K = 2^p$, $p \in {\Bbb N}$.} \label{fig:sens3} \end{figure}

We have considered binary pulsars in circular orbits in Sec.
\ref{sec:ps}. In Sec. \ref{sec:fa}, we derived the analytical solution
to the problem of Fourier analysis of such pulsar signals in the solar
system barycentric frame, in the continuous and discrete regimes. We
found that each harmonic in the power spectrum of such signals spreads
over about $2\,\,N_{P_{\rm orb}}\,\varphi_n$ bins, within which it is
$N_{P_{\rm orb}}$ periodic. We suggested two ways of modelling the
Fourier response of such pulsar signals and estimated the computational
cost of these methods. In Sec. \ref{sec:pst}, we have estimated the
detectability of such pulsars (assuming that all parameters were
unknown) using three widely used methods (Method A : standard Fourier
analysis - Method B : constant acceleration correction - Method C :
stack searches) for which we derived efficiency factors $\gamma$. In
Sec. \ref{sec:pcrt}, we presented a detailed description of our Partial
Coherence Recovery Technique (Method D) which takes advantage of the
periodicity in the power spectrum of a binary pulsar signal (see Sec.
\ref{sec:fa}) to implement a pulsar search algorithm based on the Phase
Modulation Search method (Ransom 1999). In this section, we present and
discuss the sensitivity of all four techniques to close binary pulsars
of different types with various companion masses and orbital periods.

\begin{figure} \resizebox{8.6cm}{7.1cm}{\includegraphics{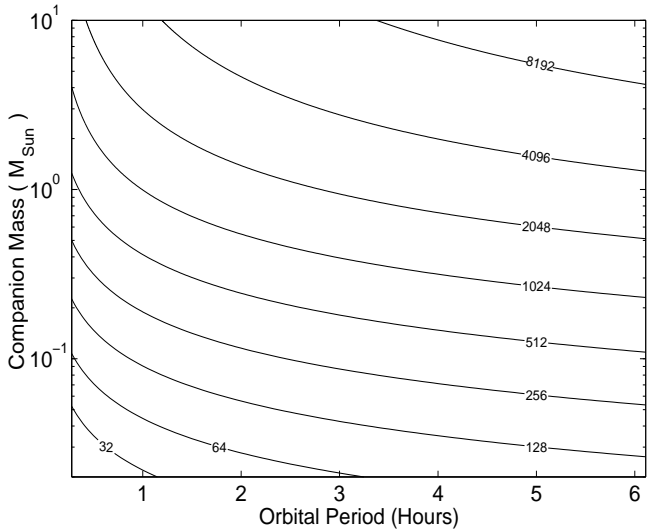}}
\caption[]{Contour map of $\varphi_1$ as a function of orbital period
$P_{\rm orb}$ and companion mass M$_2$. We assumed $i=90^o$ and
$\nu_{\rm psr}=250$ Hz. We recall that $\varphi_1 \propto \nu_{\rm
psr}\,\sin(i)$.} \label{fig:binaries} \end{figure}

\noindent We have simulated $2^{17}$ random binary systems with

\mathindent 2cm \[\indent\left\{\begin{array}{ll} \varphi_1 & \in
[1;10000]  \\ N_{P_{\rm orb}} & \in [1.5;20] \\ \phi_{orb} & \in
[-\pi;\pi] \\ \eta & \in [-0.5;0.5] \end{array}\right.\]

\noindent and calculated $\gamma_{\rm D}$ for each system individually.
We compare our results with Method A, B and C on Fig. \ref{fig:sens1},
\ref{fig:sens2} and \ref{fig:sens3}. We recall that the sensitivity of
pulsar search algorithms usually expressed in units of milli-Jansky is
inversely proportional to $\gamma$.

\noindent As shown in Fig. \ref{fig:sens1}, $\gamma_{\rm D}/\gamma_{\rm
A} \gsim 3$ for all $\varphi_n$ and both methods are strongly
correlated with $\varphi_n$. However, $\gamma_{\rm D}$ varies relatively
little with $N_{P_{\rm orb}}$ as we can see in Fig. \ref{fig:sens2}.
Fig. \ref{fig:sens2} and \ref{fig:sens3} compare state-of-the-art
methods in recovering smeared pulsar signals. Method D is significantly
better than Method B when searching for very close binary systems with
ultra-short orbital periods for which $N_{P_{\rm orb}}$ is large (e.g.
$N_{P_{\rm orb}}=36$ for $P_{\rm orb}=10$ minutes and $T_{\rm obs}=6$
hours) and relatively small $\varphi_n$. Furthermore, $\gamma_{\rm B}$
in Fig. \ref{fig:sens2} represents an ideal case where it is possible to
run a constant acceleration code on about one seventh of $P_{\rm orb}$,
which is rarely the case as the orbital period is usually unknown.
Therefore, we expect $\gamma_{\rm D}/\gamma_{\rm B}$ to be even larger.
Another important consideration concerns the processing time. In Method
B many long FFTs are taken repeatedly for each positive and negative
acceleration trial for various data lengths, whereas in Method D only
one long FFT is performed. The gain in processing time is considerable,
especially when the range of orbital accelerations is large. Method D
(PCRT) does not perform well when $N_{P_{\rm orb}} < 2$ because a
detection then relies on an aliased frequency component (Shannon 1948).

\noindent Fig. \ref{fig:sens3} demonstrates that Method D is always more
sensitive than Method C. The
reason is as follows : as seen in Sec. \ref{sec:fa}, a harmonic
manifests itself in ${\cal P}$-space as a comb of sidebands separated by
$N_{P_{\rm orb}}-1$ bins. These bins carry mostly noise and are co-added
de facto in Method C. Whereas in Method D, we perform a spectral
analysis which basically ignores them.  However, Method C must be
regarded as a useful method when compared to other standard searching
techniques. Its good performance resides in the distribution of the
noise itself, which becomes gaussian because many values are co-added,
while, due to the decrease in frequency resolution, one deals with many
fewer points hence improving the significance level of a detection. Such
a method is also computationally very efficient. A clear drawback
however resides in its lack of information about the binary system and
poor frequency resolution.

\noindent In order to estimate the detectability of pulsars in
ultra-short binary systems, we have plotted $\varphi_1$ as a function of
$P_{\rm orb}$ and companion mass M$_2$ for a 4 ms pulsar in Fig.
\ref{fig:binaries}. This plot can be used for determining $\varphi_1$
for any particular pulsar. For example, let us consider a 8 ms pulsar in
a binary system with $P_{\rm orb} = 1$ hour and M$_2 = 0.1$ M$_{\odot}$.
The value of $\varphi_1$ is $128*4/8 = 64$ radians, where $4$
corresponds to the reference pulsar period in Fig. \ref{fig:binaries}
and $8$ corresponds to the pulsar period we consider in milliseconds. Of
course, $\varphi_1$ is a maximum value since we assumed an inclination
angle of $90$ degrees but we recall that $\varphi_n$ scales directly
with $\sin(i)$. Let us assume that $T_{\rm obs} = 6$ hours, hence
$N_{P_{\rm orb}}=6$. Looking up the $\gamma$ values for Method A, B, C
and D in Fig. \ref{fig:sens1}, \ref{fig:sens2} and \ref{fig:sens3}, we
see that this particular binary system would be best detected using
Method D (PCRT) as $\gamma_{\rm D} > \gamma_{\rm C} > \gamma_{\rm A} >
\gamma_{\rm B}$  (see Sec. \ref{sec:pst} for an accurate definition of
$\gamma_{\rm A} $ and $\gamma_{\rm B}$).

\noindent Of particular interest is the detection of pulsars with large
companion masses such as Neutron Star - Neutron Star and Neutron Star -
Black Hole systems in circular orbits. In such systems, millisecond
pulsar signals are smeared over many bins in the Fourier spectrum
because $\varphi_n$ is large. These systems are characterized by large
accelerations which are usually not searched for in constant
acceleration searches due to the large increase in the number of
acceleration trials, which tremendously increases the processing time.
Stack searches could be very useful in such a context. However, the PCRT
is, on average, the most efficient and sensitive method available for
detecting such extreme systems, with the proviso that the orbital period
is short so that $N_{P_{\rm orb}}$ is greater than~2.

\noindent By direct and detailed comparison with other methods for
correcting for the period variation caused by orbital acceleration, we
find the PCRT is a very efficient method for detecting fast rotating
radio/X-ray pulsars in binary systems, especially when $N_{P_{\rm orb}}$
is large. However, it requires long observations, and therefore is not
well suited for direct application to all-sky pulsar surveys where
integration times are usually less than half an hour. On the other hand,
the method is ideally suited to long observations of globular and open
clusters, steep spectrum point sources and X-ray point sources.

\appendix

\onecolumn

\section{} \label{appendix:table}

Main symbols defined in the text, with description and section of first
use.

\begin{table*}[h!] \centering \begin{tabular}{@{}lll@{}} Symbol &
Description & Section \\ [0.1cm] \hline \\ $\chi^{D_f}$ 	&   Inverse
cumulative density function for a chi-squared \\ &	distribution with
$D_f$ degrees of freedom. & \ref{sec:pst} \\ [0.1cm] $\gamma$		&
Efficiency of a binary pulsar search method ($\,0.0 < \gamma < 1.0\,$).
& \ref{sec:pst} \\  [0.1cm] $H$			&	Harmonic fold number in
${\cal Q}$-space.

& \ref{sec:dl} \\ [0.1cm] $I(\varphi_{n},N_{P_{\rm orb}})$ & 	Width
of $\tilde{\zeta}_n$ in bins (about $2\,\,N_{P_{\rm orb}}\,\varphi_n$
bins). & \ref{sec:fs} \\ [0.1cm] $K$			&	Length of a sliding
window (${\cal W}_K$) in bins.

& \ref{sec:ccost}, \ref{sec:detect} \\ [0.1cm]
$L$			&	Number of spectral samples searched for a pulsar
signal in ${\cal P}$-space. & \ref{sec:sw} \\  [0.1cm] $\eta$      	&
Spectral leakage factor~: $\eta \in [-0.5;0.5]$ .

& \ref{sec:ccost} \\ [0.1cm] $N$ 			& 	Number of equally
spaced sampling points in T-space. & \ref{sec:fs} \\ [0.1cm]
$N_A$			&	Number of possible harmonic fold combinations. &
\ref{sec:dl} \\ [0.1cm] $N_H$			&	Number of samples which
can be folded exactly $H$ times. & \ref{sec:dl} \\  [0.1cm]
$N_K$			& 	Number of sliding windows ${\cal W}_{K}$ used in
the analysis. & \ref{sec:sfft} \\ [0.1cm]

$N_{\rm s}$	& 	Number of sub-segments used in Method C.

& \ref{sec:pst} \\ [0.1cm]

$P_{\zeta,n}$	&	Power spectrum of $\tilde{\zeta}_n$ ($\sim
I(\varphi_{n},N_{P_{\rm orb}})$ bins wide). & \ref{sec:fs} \\ [0.1cm]
${\cal P}_n$	&	Noisy $P_{\zeta,n}$. & \ref{sec:anc} \\  [0.1cm]
${\cal Q}$	& 	Power spectrum of ${\cal P}_n$.

& \ref{sec:pcrt} \\ [0.1cm] $\varrho$		&	Percentage of overlap
between consecutive ${\cal W}_K$. & \ref{sec:sw} \\ [0.1cm]
$\rho$		& 	Number of overlapping bins between consecutive
${\cal W}_K$. & \ref{sec:sw} \\  [0.1cm] $\tilde{\zeta}_n$ &	Discrete
Fourier response of the $n^{\rm th}$ harmonic of a noiseless binary
pulsar signal \\ &	of constant amplitude, shifted back to the origin
of the frequency scale.

& \ref{sec:fs} \\ [0.1cm] $W$			&	Observing window e.g.
Boxcar ($W_{\rm B}$).

& \ref{sec:cfr} \\ [0.1cm] ${\cal W_K}$ 	&  	Sliding window of
length $K$. & \ref{sec:detect} \\ [0.1cm] $\varphi_n$	& 	Half the
phase rotation experienced by the $n^{\rm th}$ harmonic of the pulsar
signal \\ &	during the orbit, in radians~: $\varphi_{n} = 2\pi n
\nu_{\rm psr}\left[a_1\sin(i) /{\rm c}\right]$. & \ref{sec:ps} \\
[0.1cm] $\phi_{\rm orb}$ & 	Initial orbital phase in radians,
measured from superior conjunction of the pulsar. & \ref{sec:ps} \\
[0.1cm] $\nu_{\rm psr}$ , $N_{P_{psr}}$ & 	Observed pulsar rotation
frequency in Hz and Fourier bins, respectively. & \ref{sec:ps},
\ref{sec:fs} \\ [0.1cm] $\nu_{\rm orb}$ , $N_{P_{\rm orb}}$ & Observed
pulsar orbital frequency in Hz and Fourier bins, respectively. &
\ref{sec:ps}, \ref{sec:fs} \\ \\ \end{tabular} \end{table*}

\section{} \label{appendix:serie}

\mathindent 3cm

We derive the mathematical proof of Eq. (\ref{eq:mathserie}). We shall
demonstrate that

\begin{equation}
e^{\,jz\cos\theta}=\sum_{k=-\infty}^{\infty}j^{k}\,J_{k}(z)\,e^{\,jk\
theta}\,. \end{equation}

\noindent By definition,

\begin{equation} e^{\,jz\cos\theta}=\cos(z\cos\theta)+j\sin(z\cos\theta)
\label{eq:def} \end{equation}

\noindent where, according to Eqs. (9.1.44) and (9.1.45) in 
Abramowitz \& Stegun 1974, 

\begin{equation}
\cos(z\cos\theta)=J_{0}(z)+2\,\sum_{p=1}^{\infty}(-1)^{p}\,J_{2p}(z)\,\
cos(2p\,\theta) \label{eq:cos} \end{equation}

\begin{equation}
\sin(z\cos\theta)=2\,\sum_{p=0}^{\infty}(-1)^{p}\,J_{2p+1}(z)\,\cos(\,[
2p+1]\,\theta) \label{eq:sin}\,. \end{equation}

\noindent Since $j=(-1)^{1/2}$, Eq. (\ref{eq:sin}) can be rewritten as

\begin{equation}
\sin(z\cos\theta)=\frac{2}{j}\,\sum_{q=1/2}^{\infty}(-1)^{q}\,J_{2q}(z)\
,\cos(2q\,\theta) \label{eq:sin2} \end{equation}

\noindent where $q=p+\frac{1}{2}$. Combining Eqs. (\ref{eq:cos}) and
(\ref{eq:sin2}) in (\ref{eq:def}) yields

\begin{equation}
e^{\,jz\cos\theta}=J_{0}(z)+2\,\sum_{k=1}^{\infty}(-1)^{k/2}\,J_{k}(z)\,
\cos(k\,\theta)\,. \end{equation}

\noindent Since $\cos(k\,\theta)= \left[
e^{\,jk\,\theta}+e^{-jk\,\theta} \right]/2$, we get

\begin{equation} e^{\,jz\cos\theta} =  J_{0}(z) + \sum_{k=1}^{\infty}
\left[
\,j^{k}\,J_{k}(z)\,e^{\,jk\,\theta}+j^{k}\,J_{k}(z)\,e^{\,-jk\,\theta}
\,\right]\,. \end{equation}

\noindent According to Eq. (9.1.5) in Abramowitz \& Stegun 1974 : 
$J_{k}(z)=j^{-2k}\,J_{-k}(z)$. Therefore,

\begin{equation} e^{\,jz\cos\theta} = J_{0}(z)+\sum_{k=1}^{\infty}\left[
\,j^{k}\,J_{k}(z)\,e^{\,jk\,\theta}+j^{-k}\,J_{-k}(z)\,e^{\,-jk\,\theta}
\,\right] \end{equation}

\noindent which simplifies to

\begin{equation}
e^{\,jz\cos\theta}=\sum_{k=-\infty}^{\infty}j^{k}\,J_{k}(z)\,e^{\,jk\,\
theta}\,. \end{equation}

\noindent We note that in Eq. (\ref{eq:mathserie}), $j^{k}$ has been
replaced by its equivalent exponential notation $e^{\,jk\pi/2}$.

\section{} \label{appendix:offt}

\subsection{Determination of Eq. (\ref{eq:offt1})}
\label{appendix:offt1}

Let us consider a given frequency channel and compute $N_K$ FFTs of
length $K \in [K_{\rm min} \ldots K_{\rm max}]$. The number of
operations required for an FFT of length $K$ is given by $K \log_2 K$.
Therefore,

\begin{equation} {\cal O}_{\rm FFT} = K_{\rm min} \log_2 K_{\rm min} +
\ldots + [\,2^{N_K-1} K_{\rm min}\,]\,\log_2\,[\,2^{N_K-1} K_{\rm
min}\,] \end{equation}

\noindent which can be rewritten as

\begin{equation} {\cal O}_{\rm FFT} = K_{\rm min} \left[\,\log_2 (K_{\rm
min}) \sum_{k=0}^{N_K-1} 2^k + \sum_{k=0}^{N_K-1}  2^k k\,\right]\,.
\end{equation}

\noindent The summations can be explicitly computed so that

\begin{equation} {\cal O}_{\rm FFT} = K_{\rm min} \left[ \, \log_2
(K_{\rm min}) \, [\,2^{N_K}-1\,] + [\,2^{N_K}(N_K-2)+2\,] \, \right]\,.
\label{eq:o11} \end{equation}

\noindent Assuming $2^{N_K} \gg 1$, Eq. (\ref{eq:o11}) subsequently
reduces to

\begin{equation} {\cal O}_{\rm FFT} \simeq 2^{N_K} K_{\rm min}
\left[\,\log_2 (K_{\rm min}) + N_K - 2\,\right]\,. \end{equation}

\noindent Naturally, ${\cal O}_{\rm FFT}$ must be multiplied by $L$
hence Eq. (\ref{eq:offt1}).

\subsection{Determination of Eq. (\ref{eq:offt2})}
\label{appendix:offt2}

We now consider $\rho > 1$. The number of frequency channels at which a
FFT of length $K$ is evaluated reduces to $L/K(1-\varrho)$. Therefore,

\begin{equation} {\cal O}_{\rm FFT} = \frac{L\,\left[\,K_{\rm min}
\log_2 K_{\rm min} \,\right]}{K_{\rm min} (1-\varrho)} + \ldots +
\frac{L\,\left[\,2^{N_K-1} K_{\rm min} \log_2 (2^{N_K-1} K_{\rm min})
\,\right]}{2^{N_K-1} K_{\rm min} (1-\varrho)}\,. \end{equation}

This equation can be rewritten in a simple form as

\begin{equation} {\cal O}_{\rm FFT}  = \frac{L}{1-\varrho}\,\left[\,N_K
\log_2 K_{\rm min} + \sum_{k=0}^{N_K-1} k\,\right]\,. \end{equation}

Subsequently, we obtain Eq. (\ref{eq:offt2})

\begin{equation} {\cal O}_{\rm FFT}  = \frac{L}{1-\varrho}\,\left[\,N_K
\log_2 K_{\rm min} + \frac{N_K (N_K-1)}{2}\,\right]\,. \end{equation}

\twocolumn


\end{document}